\documentclass[12 pt]{article}
\usepackage{times}
\usepackage{subfigure}

\usepackage{pifont}
\usepackage{floatrow}
\usepackage{booktabs}

\usepackage{caption}
\usepackage{mathrsfs}
\usepackage[fleqn]{amsmath}
\usepackage{amsfonts,amsthm,amssymb,mathrsfs,bbding}
\usepackage{txfonts}
\usepackage{graphics,multicol}
\usepackage{graphicx}

\usepackage{epstopdf}

\usepackage{graphicx}
\usepackage{booktabs}
\usepackage{xcolor}
\usepackage{amsmath}

\usepackage{color}
\usepackage{caption}
\usepackage{indentfirst}
\usepackage{cite}
\usepackage{latexsym,bm}
\usepackage{enumerate}
\usepackage{epsfig}
\usepackage[noend]{algpseudocode}
 \usepackage[ruled,vlined]{algorithm2e}

\evensidemargin=\oddsidemargin\topmargin=-1.5cm

\theoremstyle{definition}

\newtheorem{remark}{Remark}

\addtocounter{section}{0}

\def\x{\bm{x}}

\def\y{\bm{y}}

\def\v{\bm{v}}
\def\b{\bm{b}}
\def\s{\bm{s}}
\def\@#1{{\cal #1}}

\def\dim{d}
\def\Beta{\bm{\beta}}

 \newcommand{\Cov}{\operatorname{Cov}}

\newcommand{\argmin}{ \operatornamewithlimits{argmin} }

\numberwithin{equation}{section}

\let\trueiiint=\iiint
\def\iiint{\mathop{\textstyle\trueiiint}\limits}
\def\intinfty{\int\limits_{\!\!-\infty\,\,}^{\,\,\infty\!\!}\kern-0.0em}
\def\iintinfty{\mathop{\int\!\!\int}\limits_{\!\!-\infty\,\,}^{\,\,\infty\!\!}\kern-0.0em}
\def\iiintinfty{\mathop{\int\!\!\int\!\!\int}\limits_{\!\!-\infty\,\,}^{\,\,\infty\!\!}\kern-0.0em}

\let\<=\langle
\let\>=\rangle
\let\^=\hat
\def\~#1{{\-ox{\sf#1}}}

\let\-=\mathbf

\def\@#1{{\cal #1}}

\def\Cov{\mathrm{Cov}}

\let\<=\langle
\let\>=\rangle

\let\^=\hat

\def\circ{\ifmmode\mathchar"220E\else$\mathchar"220E$\fi}
\def\@#1{{\cal #1}}
\def\x{{\bm{x}}}
\def\f{{\bm{f}}}

\def\s{\bm{s}}

\begin{document}
\title{Model-Embedded Gaussian Process Regression for Parameter Estimation in Dynamical System}

\author{Ying Zhou$^{a}$ Jinglai Li$^{b}$ Xiang Zhou$^{c}$ and  Hongqiao Wang$^{d *}$\\
{\it $^{a}$Business School} \\
{\it Central South University} \\
{\it Changsha 410083, P.R. China}\\
\ \\
{\it $^{b}$School of Mathematicsl} \\
{\it University of Birmingham} \\
{\it Edgbaston, Birmingham, B15 2TT, UK}\\
\ \\
{\it $^{c}$School of Data Science and Department of Mathematics} \\
{\it City University of Hong Kong} \\
{\it Tat Chee Ave., Kowloon, Hong Kong SAR, P.R. China}\\
\ \\
{\it $^{d}$School of Mathematics and Statistics} \\
{\it Central South University} \\
{\it Changsha 410083, P.R. China}\\
[2mm]
{$^*$ Corresponding author: Hongqiao Wang}\\
{\it School of Mathematics and Statistics} \\
{\it Central South University} \\
{\it Changsha 410083, People’s Republic of China}\\
{\it E-mail: hongqiao.wang@csu.edu.cn}}

%


\date{}

\maketitle {\flushleft\large\bf Abstract }
Identifying dynamical system (DS) is a vital task in science and engineering. Traditional methods require numerous calls to the DS solver, rendering likelihood-based or least-squares inference frameworks impractical.
For efficient parameter inference, two state-of-the-art techniques are the kernel method for modeling and the "one-step framework" for jointly inferring unknown parameters and hyperparameters. The kernel method is a quick and straightforward technique, but it cannot estimate solutions and their derivatives, which must strictly adhere to physical laws.
We propose a model-embedded "one-step" Bayesian framework for joint inference of unknown parameters and hyperparameters by maximizing the marginal likelihood. This approach models the solution and its derivatives using Gaussian process regression (GPR), taking into account smoothness and continuity properties, and treats differential equations as constraints that can be naturally integrated into the Bayesian framework in the linear case.
Additionally, we prove the convergence of the model-embedded Gaussian process regression (ME-GPR) for theoretical development. Motivated by Taylor expansion, we introduce a piecewise first-order linearization strategy to handle nonlinear dynamic systems. We derive estimates and confidence intervals, demonstrating that they exhibit low bias and good coverage properties for both simulated models and real data.

\begin{flushleft}
\end{flushleft}


%
%

\section{Introduction}
Dynamical system, represented as a set of ordinary differential equations (ODEs), are commonly used to model behaviors in scientific domains, such as gene regulation\cite{hirata2002oscillatory}, biological rhythms\cite{forger2017biological}, spread of disease\cite{miao2009differential}, ecology\cite{busenberg2012differential}, etc.
In consequence, they are often studied by statisticians, who are interested particularly in estimating model parameters and determining any smoothing, or tuning, parameters that are needed to estimate the model.
Dynamical systems model output change directly by linking the output derivatives $\dot{\x(t)}$ to $\x(t)$ itself:
\begin{equation}
\label{eq:dynamic_system}
    \dot{\x}(t) \equiv \frac{d\x(t)}{dt}= \f(\x(t), t;\Theta), \quad t\in [0, t_{max}],
\end{equation}
where $\x = [x_1,\dots,x_D]^\top$ denote an $D$-dimensional function of a scalar variable $t$, $\dot{\x}$ denotes the first order derivatives of $\x$, dynamic model $\f (\x(t), t;\Theta)\in \mathbb{R}^D$ is a column vector consisting of the basis function $f_i$ evaluated at $(\x,t,\Theta)$ and $\Theta = \{\theta_1,\dots,\theta_p\}$ contains $p$ model parameters defining the system whose values are not known from experimental data, theoretical considerations or other sources of information.
Systems involving derivatives of $x_i$ of order $M > 1$ are reducible to expression \eqref{eq:dynamic_system} by defining new variables, $x_{i+1} = \dot{x}_i$ and $x_{i+2} = \dot{x}_{i+1},\dots,x_{i+M} = \dot{x}_{i+M-1}$.
When $\f$ is nonlinear, solving $\x(t)$ given initial conditions $\x(0)$ and $\Theta$ generally requires a numerical integration method, such as Runge-Kutta, which would be computational intensive.

In practice, we can only directly observe the noisy observation  $\y$ instead of $\x$. We assume that $\x$ is observed with measurement error and specifically, the noisy measurements satisfy 
\begin{equation} 
\label{y}
\y = \x + \bm{\epsilon},
\end{equation}
where $\bm{\epsilon}$ is the measurement error, also called noisy/observation error. Assumption is made that $\bm{\epsilon}$ is a vector of  {\it iid} random variables and follows the multivariate Gaussian distributions with mean $\bm{0}$ and covariance $\text{diag}(\bm{\sigma})$, where $\bm{\sigma} =  [\sigma_1^2,\dots,\sigma_D^2]^\top$.
The observed data are stored in the data-set $\mathcal{D} = \{T,Y\}$, where $T = [t_1,\dots,t_n]$ and $Y=[\y_1,\dots,\y_n]$, means that $D$-dimensional observations $\y_i\in \mathbb{R}^D$ can be collected at time index $t_i$.
In practical scenarios, the presence of unobserved system components poses a considerable challenge.

Point estimation and posterior inference  are usually used in dynamical systems for the parameter estimation.
Point estimation  for the model parameter in dynamical systems involves the determination of the "optimal" estimate by maximizing either the likelihood function $p(\mathbf{y}|\hat{\mathbf{x}}(t,\Theta), \bm{\sigma})$ or the posterior distribution $p(\Theta|\mathbf{y})\propto p(\mathbf{y}|\hat{\mathbf{x}}(t,\Theta), \bm{\sigma}) p(\Theta)$, where $\hat{\mathbf{x}}(t,\Theta)$ represents the numerical solution of the DS obtained through numerical integration with the given parameter set $\Theta$ and initial conditions.
On the other hand, posterior inference  for parameter estimation in dynamical systems, is focused on deriving the posterior distribution, which encapsulates the uncertainty information associated with the estimation process.
Nevertheless, practical implementation of methods that directly infer or maximize the likelihood or posterior distribution encounters a significant computational challenge. This is due to the necessity of performing numerical integration for each sampled $\Theta$ within optimization or Markov chain Monte Carlo (MCMC) procedures, as highlighted by Calderhead and Girolami \cite{calderhead2008accelerating}.

Indeed, the high computational cost of repeatedly solving differential equations has prompted statisticians to seek alternative approaches to this problem, using non-parametric methods. 
The best known technique is parameter cascade \cite{ramsay2007parameter}, where an innovative penalized likelihood approach uses a B-spline basis for constructing estimated functions to simultaneously satisfy the ODE system and fit the observed data.
The parameter cascade method balances these two conflicting requirements by minimizing the weighted average of two respective criterion functions.
If the substantive manual input like the spline order, the number of knots, the balance weights, etc, are all predecided, the  estimation of the model parameters can be obtained together with the spline coefficients in a single optimization which  is referred to as a joint estimation procedure, namely "one-step" method.
With the success of parameter cascade method,  \cite{cao2008estimating,hooker2009forcing,hooker2011parameterizing} explore it in various applications.
Encouraged by the success of spline-based non-parametric methods in these problems, kernel-based non-parametric methods were explored by Liang \cite{liang2008parameter}, Jake Bouvrie  \cite{doi:10.1137/14096815X} and Houman Owhadi \cite{HAMZI2023133853, hamzi2021learning}.
Specifically, Houman Owhadi proposed a Kernel-Flows method for selecting or learning a "best" kernel for dynamical systems.
The basic idea of Liang \cite{liang2008parameter} was, first, to estimate the regression function and its derivative by using kernel-based non-parametric methods, e.g. techniques founded on local polynomials, and then to adjust the parameter values so that the non-parametric estimators satisfied most closedly the differential equation requirement.
This kernel-based methods perform non-parametric estimation while completely ignoring the model, and then derive the model parameters by minimizing the model discrepancy expressed in terms of the estimated functions.
Since the non-parametric regression and parameter estimation stages are performed separately, these methods are referred to as two-step estimation procedures.

As an alternative to the modeling tool, Gaussian processes (GPs) are a nature candidate for fulfilling the smoothing role in a Bayesian paradigm due to their flexibility and analytic tractability\cite{williams2006gaussian}.
The use of GPs to approximate the dynamical system and facilitate computation has been previously studied by a number of authors\cite{yang2021inference,dondelinger2013ode,wenk2019fast,calderhead2008accelerating,barber2014gaussian,ghosh2017fast,lazarus2018multiphase}.
The basic idea is to specify a joint GP over $\y$, $\x$,$\dot{\x}$ with hyperparameters $\phi$ and then, provide a factorization of the joint density $p(\y,\x,\dot{\x}, \Theta,\phi,\sigma)$ that is suitable for inference.
The main challenge is to find a coherent way to combine information from two distinct sources: the approximation to the system by the GPs governed by hyperparameters $\phi$ and the actual dynamical system equations governed by parameters $\Theta$.
References \cite{calderhead2008accelerating,dondelinger2013ode} use a product of experts heuristic by letting $p(\dot{\x}|\x,\Theta,\phi)\propto p(\dot{\x}|\x,\phi)p(\dot{\x}|\x,\Theta)$, where the two distributions in the product come from the GPs and a noise version of the ODE, respectively.
A graphical model is used in \cite{wenk2019fast} instead of the product of experts heuristic with an artificial noisy version of the ODE.
In \cite{barber2014gaussian},  the authors start with a different factorization: $p(\y,\x,\dot{\x},\Theta,\phi,\sigma) = p(\y|\dot{\x},\phi,\sigma)p(\dot{\x}|\x,\Theta)p(\x|\phi)p(\Theta)$, where $p(\y|\dot{\x},\phi,\sigma)$ and $p(\x|\phi)$ are given by the GP and $p(\dot{\x}|\x,\Theta)$ is a Dirac delta distribution given by the ODE.

Inspired by the success of kernel-based Gaussian Process (GP) methods and existing "one-step" approaches, this paper introduces a competitive method that estimates model parameters and tuning parameters simultaneously without numerical integration. By reformulating dynamical systems in Section \ref{set:LDS}, all components of the system's solution are modeled as Gaussian processes in Section \ref{set:Gaussian_process}.
We construct a joint Bayesian inference framework that incorporates observations and constraints of differential equations, encompassing model parameters, dynamical system solutions, and hyperparameters in Section \ref{set:ME-GP}. A semi-stochastic gradient descent (Semi-SGD) method is used to optimize the marginal likelihood with finite data while satisfying the constraints of equation information over an infinite time interval.
For nonlinear dynamical systems, Taylor expansion and piecewise linearization are applied in Section \ref{set:linearization}.  Three numerical examples are provided to illustrate the effectiveness of the proposed method in Section \ref{set:num_examples}.

\section{Latent Variable Reformulation of Linear Dynamical Systems}
\label{set:LDS}

Our goal is to estimate the unknown parameters $\Theta$ based on the given noisy dataset $\mathcal{D}$ and the specific form of differential equations (containing unknown parameters).
The estimation problem in linear dynamic process where $\f$ in  Equation \eqref{eq:dynamic_system} is a linear operator or a combination of linear operators w.r.t. $\x$ is more fundamental and has been serving as the very start of nonlinear dynamical systems.
Then we firstly consider the problem of estimating the unknown model parameters in the linear dynamical system, i.e. each $f_i$ in Equation \eqref{eq:dynamic_system} is linear operator.

The linear dynamical system can be formulated as:
\begin{equation}
\label{eq:linear_ds}
\begin{bmatrix}
\dot{x}_1\\
\vdots\\
\dot{x}_D
\end{bmatrix}
=J\cdot\begin{bmatrix}
x_1\\
\vdots\\
x_D
\end{bmatrix} + \bm{b}
\end{equation}
where $J$ is a $D\times D$ coefficient matrix and $\bm{b}\in \mathbb{R}^D$.

We focus on a particular class of dynamical systems, where Equation \eqref{eq:linear_ds} can be equivalently transformed into the form of Equation \eqref{eq:linear_dynamic_g}. In this context, we denote $u \equiv x_k$, representing certain specific component of the solution vector $\x = [x_1, \dots, x_D]$.  For clarity and consistency, $u$ and $x_k$ will be used interchangeably throughout the paper.
The terms corresponding to each dimension of the solution may be reformulated as functions of the variable $u$ and its derivatives $\frac{d^m u}{dt^m}$, as outlined below:

\begin{equation}
\label{eq:linear_dynamic_g}
\begin{bmatrix}
x_1\\
\vdots\\
x_D
\end{bmatrix}
=G\cdot \begin{bmatrix}
\frac{du}{dt}\\
\vdots\\
\frac{d^Mu}{dt^M}
\end{bmatrix}+ \bm{\nu} 
\end{equation}
where $G = [g_{ji}], j=1,\dots,D, i=1,\dots,M$ and $\bm{\nu} =[\nu_1,\dots,\nu_D]^\top\in \mathbb{R}^D$. 
$g_{ji}$ and $\nu_j$ are the functions of the elements of $J$ and $\bm{b}$. 
It is worth emphasizing that the method introduced in this paper is not applicable to the general dynamical system frameworks in Equation \eqref{eq:linear_ds}.
However, this transformation ( from \eqref{eq:linear_ds} to \eqref{eq:linear_dynamic_g}) is frequently encountered in various dynamical systems, including the Susceptible-Infected-Removed (SIR) epidemic model and the FitzHugh-Nagumo systems, among others.
The mathematical representation given by \eqref{eq:linear_dynamic_g} provides a comprehensive understanding of the variable 
$\x$. This perspective defines 
$\x$ as a set of interrelated Gaussian processes, with the correlation governed by the system's inherent dynamics and the unknown parameters denoted by $\Theta$.
Details would be discussed in Section \ref{set:ME-GP}.

In the following section, we give a brief introduction of Gaussian process and highlight a key fundamental property of Gaussian process which contribute the modeling of correlative Gaussian process.

\section{ Gaussian process}
\label{set:Gaussian_process}
\subsection{Prior}

Given a one-dimensional ordinary differential equation solution $x(t)$, the GP method constructs a surrogate model of $x(t)$ in a nonparameteric Bayesian regression framework. Specifically the target function $x(t)$ is cast as a Gaussian random process whose mean $\mathbb{E}[x(t)]\equiv 0$ and covariance is specified by a kernel function $\text{COV}[x(t),x(t')] = k_{xx}(t,t')$,
which is denoted as 
\begin{equation}
\label{eq:u_gp}
x(t)\sim \mathcal{GP}(0, k_{xx}(t,t';\beta_\alpha, \beta_l)),
\end{equation}
where $\beta_\alpha$ and $\beta_l$ denote the hyper-parameters of the kernel function $k_{xx}$, defined in Equation \eqref{eq:kerenl_uu}. 
The kernel $k_{xx}$ allows us to encode any prior knowledge we may have about $x(t)$, and can accommodate the approximation of arbitrarily complex functions. 
The choice of the specific form of $k_{xx}$ will be discussed later in Section \ref{set:kernel}.

The key property of Gaussian process in our favor is that any linear transformation, such as differentiation, integration and the combination of linear operators, of a Gaussian process is still a Gaussian process. 
With the assumption \eqref{eq:u_gp},  we consider the
linear operator, $ \mathcal{L} $, acted on  $x(t)$.
Then $w = \mathcal{L} x(t)$ is also a mean-zero Gaussian process 
\begin{equation}
w \sim \mathcal{GP}(0, k_{ww}(t,t')) 
\end{equation}
where {$k_{ww}(t,t')=\Cov(\mathcal{L}x(t), \mathcal{L}x(t'))$} denotes the covariance function  of   $w$ between    $t$ and    $t'$.
 The following fundamental relationship between the kernels $k_{xx}$ and $k_{ww}$ is well-known
(see  e.g. \cite{seeger2004gaussian,zhou2022inferring,wang2021explicit}),
\begin{equation}
\label{eq:k_GG}
k_{ww}(t,t';\beta_\alpha, \beta_l) = \mathcal{L}_t \mathcal{L}_{t'} k_{xx}(t,t';\beta_\alpha, \beta_l).
\end{equation}
Here  we add the subindex in  the linear differential operator $ \mathcal{L} $  to specify  the differentiation is for $t$ or ${t}'$ variable
in the kernel function.
$k_{xx}$ and $k_{ww}$ share the same hyper-parameter $\beta_\alpha, \beta_l$.
Simiarly, 
for the covariance between   $x$ and   $w$,
$k_{xw}(t,t')=\Cov(x(t), \mathcal{L}x(t'))
$ and $k_{wx}(t,t')  = \Cov(\mathcal{L}x(t), x(t'))$, we have 
\begin{equation}
\label{eq:k_uw}
k_{xw}(t,t';\beta_\alpha, \beta_l)= \mathcal{L}_{t'}k_{xx}(t,t';\beta_\alpha, \beta_l),
~\text{ and } 
k_{wx}(t,t';\beta_\alpha, \beta_l) = \mathcal{L}_{t}k_{xx}(t,t';\beta_\alpha, \beta_l).
\end{equation}
The specific expression of the kernels in \eqref{eq:k_uw} would be discussed in Section \ref{set:kernel}.

\subsection{Kernel}
\label{set:kernel}
The kernel (covariance function) is the crucial ingredient in a Gaussian process predictor, as it encodes our assumptions about the function   we wish to learn. Without loss of generality, the Gaussian prior of the solution used in this work is assumed to have a squared exponential covariance function (other kinds of kernels are also suitable in this framework), i.e.,
\begin{equation}
\label{eq:kerenl_uu}
k_{xx}(t, t',\beta_\alpha, \beta_l) = \beta^2_\alpha \exp(-\frac{1}{2} \frac{(t-t')^2} {\beta_l^2})
\end{equation}
where $\beta^2_\alpha$ is a variance parameter and $\beta_l$ is the length scale parameter. 
The squared exponential covariance function chosen above implies smooth approximations. 
More complex function class can be accommodated by appropriately choosing kernels. 
For example, non-stationary kernels employing nonlinear warpings of the input space can be constructed to capture discontinuous response. In general, the choice of kernels is crucial and in many cases still remains an art that relies on one's ability to encode any prior information (such as known symmetries, invariant, etc.) into the regression scheme. 
In our problem here  related to the differential operator $\mathcal{L}$,
 we require that the kernel satisfies the regularity such that the derivatives of the kernel,
 $\frac{\partial^2}{\partial t \partial t'}k_{xx}(t,t')$
 ,  which is the covariance function  of $\dot{x}$,  are at least continuous. 
 Our choice of squared exponential covariance function 
 surely satisfies this basis requirement. 

The kernel $k_{ww}$, $k_{wx}$ and $k_{xw}$ can be easily computed based on the specific definition of $k_{xx}$ in \eqref{eq:kerenl_uu}, such as, for $w = \mathcal{L}x:=(t-1)^2 \frac{dx}{dt} +x$, the covariance function $k_{wx}$ can be computed by:
\begin{equation}
  \label{klu}
  \begin{split}
  k_{wx}(t,t';\beta_\alpha, \beta_l)&:=(t-1)^2\frac{d k_{x,x}(t,t';
  \beta_\alpha, \beta_l)}{d t} + k_{x,x}(t,t';
  \beta_\alpha, \beta_l)\\
 &= -(t-1)^2\beta_l  (t-t') k_{xx}(t,t',\beta_\alpha, \beta_l) + k_{xx}(t,t',\beta_\alpha, \beta_l).
 \end{split}
 \end{equation}
 Similarly, the specific expressions of $k_{ww}$ and $k_{xw}$ can be easily computed \cite{wang2021explicit}.

Due to  the irreducible  measure noise of $x$, in practice the covariance of prior $x$ is replaced by the summation of $k_{xx}(t,t')$ and a noise kernel $\sigma_x^2\delta_{t,t'}$, where $\delta_{t,t'}$ is Dirac delta function and the parameter of variance $\sigma_x^2$ can be optimized with the kernel parameters $\{\beta_\alpha, \beta_l\}$ together.
  In a similar style, we can also introduce the noise-level parameter $\sigma_x^2 $ for the random variable $w$ and redefine the covariance of prior $w$ as $k_{ww}(t,t') + \sigma_w^2\delta_{t,t'}$ if there has the observation of it.
  Hence, all the possible hyper-parameters in GPs are collected in a vector $\bm \beta = [\beta_\alpha, \beta_l,\sigma_x,\sigma_w]$.

\section{Model-embedded Gaussian process regression}
\label{set:ME-GP}

In the linear dynamical systems \eqref{eq:linear_dynamic_g},
each component of the solution $x_j$ can be reformulated as the functions of $u$.
For notation simplification, we denote $\x_{[- k]} = \{x_1,\dots,x_{k-1},x_{k+1},\dots,x_D\}$.
With considering the smoothness and continuity properties of the solution, we propose to model the solution $u$ by Gaussian process.  Benefit from the property of Gaussian processes, any linear transformation of a Gaussian process is still a Gaussian process,  $\x_{[- k]}$ are all Gaussian processes which derives from $u$ because each $x_j$ is defined as a combination of linear operator $g_{ij}$, acted on $u$ and its derivatives, $x_j = g_{j1}\frac{du}{dt} + g_{j2}\frac{d^2u}{dt^2} + \dots g_{jM}\frac{d^M u}{dt^M} + \nu_j$.

Based on the Gaussian process assumption of $x_k$, we set $\mathcal{L}^\Theta_k \equiv 1$ and have 
\begin{equation}
\label{eq:x_k_Gaussian}
    x_k = \mathcal{L}^\Theta_k u \equiv u \sim \mathcal{N}(0,k_{u u}(t,t',\bm{\beta})),
\end{equation}
where the subscript of $\mathcal{L}$ represents the linear operator corresponding to $g$. 
The solutions $\x_{[- k]}$ in Equation \eqref{eq:linear_dynamic_g} can also be modeled as Gaussian processes
 \begin{equation}
 \label{eq:x_j_and_r_Gaussian}
     x_j = \mathcal{L}_j^\Theta u(t) \sim \mathcal{N}(0,k_{x_j x_j}(t,t',\bm{\beta})),\quad j\ne k
 \end{equation}
where $\mathcal{L}^\Theta_j$ represents the linear operator acted on $u$ corresponding to the linear function $g_j$ with model parameters $\Theta$. 
\textbf{It is worthy to emphasize that each $k_{x_j x_j}(t,t',\bm{\beta}), j\ne k$ derives from $k_{u u}(t,t',\bm{\beta})$ in our framework which can theoretically guarantee the model constraint $x_{j} = \sum_{i=1}^M g_{ji}\frac{d^iu}{dt^i} +\nu_j$}.
This setting makes us possible to  embed $D-1$ model constraints of \eqref{eq:linear_dynamic_g} into the priors theoretically.
Then we have the carefully designed priors for all $\x$.

Specially, define a new variable $v$ based on the the $k$th equation in \eqref{eq:linear_dynamic_g}
\begin{equation}
\label{eq:uEquivx_k}
    v   = u - \sum_{i=1}^M g_{ki}\frac{d^iu}{dt^i} -\nu_k,
\end{equation}
which represents the residual of the $k$th model constraint.
Similarly, $v$ can also be modeled as a Gaussian process
 \begin{equation}
 \label{eq:x_j_and_r_Gaussian}
     v = \mathcal{L}_v^\Theta u(t)\sim \mathcal{N}(0,k_{v v}(t,t',\bm{\beta})),
 \end{equation}
where $\mathcal{L}^\Theta_v$ represents the linear operator acted on $u$ corresponding to the linear function $u - g_k(u, t;\Theta)$ with model parameters $\Theta$ and $k_{v v}(t,t',\bm{\beta})$ also derives from $k_{u u}(t,t',\bm{\beta})$.
Its covairance function expression can be obtained by Section \ref{set:Gaussian_process}. 

Therefore all the solutions $\x_{[- k]}(t)$ and the constraint variable $v(t)$ can be derived from $u$ through linear operators and are Gaussian processes.
The covariances among  $\x$ , $v$ and $u$ can be computed by Equation \eqref{eq:k_GG} and \eqref{eq:k_uw}.

\begin{remark}
It would be perfect if we can find a kernel $k_{uu}$ which satisfies the covariance function relationship $k_{uu} = k_{g_k u,g_k u}$ in theory. Then the residual variable is redundant and droppable. 
But this special kernel is difficult to be designed for a general expression of $g_k$.
\end{remark}

We refer $v$ as the {\it{model constraint}},  $v(t)\equiv 0$, and the constraint will be interpreted numerically later as the observation of zero values of the function $v(t)$ at any point $t$, in a similar way to the observation $\y_i$ of $\x$ at $t_i$.  
With the prior assumption of $u$ defined in Equation \eqref{eq:u_gp} and the kernel relationships between $\x_{[- k]}$, $v$ and $u$, we then have the priors for all $\x_{[- k]}$ and $v$ and a joint inference framework ( shown in Section \ref{set:inference} )  of multivariate Gaussian distribution for the available observation data $\y$ and the model constraint $v$ can be naturally constructed.
It is worthy noting that in this framework we theoretically guarantee the solutions satisfying $D-1$ model constraint (shown in \eqref{eq:linear_dynamic_g}) and the numerical technique for guaranteeing model constraint is applied on only one remained model constraint $v=\mathcal{L}_k^\Theta u:=0$ by a Semi-SGD method in Section \ref{set:inference}.

\subsection{Inference based on data and model constraints}
\label{set:inference}
Based on the Gaussian prior of \eqref{eq:x_k_Gaussian} and \eqref{eq:x_j_and_r_Gaussian}, we obtain the joint distribution of $Y$ and $v$
\begin{equation}
\label{eq:inference}
    \begin{bmatrix}
    Y\\
    v
    \end{bmatrix}
    \sim 
    \mathcal{N}(
    \begin{bmatrix}
    \textbf{0}\\
    0
    \end{bmatrix}
    ,
    \begin{bmatrix}
    K_{\x\x} & K_{\x v}\\
    K_{v\x} & K_{vv}
    \end{bmatrix})
\end{equation}
where $K_{\x \x}$ stores the kernel function values of
\begin{equation}
K_{\x \x}=
\begin{bmatrix}
k_{x_1 x_1}&\dots& k_{x_1 x_D}\\
\vdots &\ddots&\vdots\\
k_{x_D x_1}&\dots& k_{x_D x_D}
\end{bmatrix}
\end{equation}
at all time index and $Y = [\y_1,\dots,\y_n]\in \mathbb{R}^{D\times n \times 1}$ are the vector of all stack observations which include $D$ components of dynamical system solutions at $n$ different time indexes.
Same definitions are used in $K_{\x v}$, $K_{v\x}$ and $K_{vv}$ and detail expression of $K$ refers Appendix \ref{set:appgpr}.
We denotes the time index of our data as $T_d \in \mathbb{R}^{D\times n \times 1}$ corresponding to $Y$.
Because $v$ represents the constraint which should be satisfied at any time $t_i$ theoretically, for numerical approximation we construct a new data-set $\mathcal{D}^v = \{T_c, \bm{0}\}, T_c\in \mathbb{R}^{n_c}$ which will be introduced in Section \ref{set:samples}.
Due to the measurement noise, noise kernels $\sigma_{y_j}^2\delta_{x_i,x_j}\delta_{t,t'}$ are added in the $k_{x_i x_j}(t,t')$ with the noise level $\sigma_{y_j}^2$.
A small preset regularization kernel $\sigma_{v}^2\delta_{t,t'}$ is also added in the kernel $k_{v v}(t,t')$ for computational stability.
The log marginal likelihood function of observation $Y$ and constraint $v$ can be given as 
\begin{equation}
\label{eq:log_likelihood}
\log p([Y,v]^\top|\mathcal{D},\mathcal{D}^v,\Theta,\Beta) = -\frac{1}{2}[Y,v]
K^{-1}\begin{bmatrix}
Y\\ v
\end{bmatrix}
-
\frac{1}{2}\log |K|
-\frac{n+n_c}{2}\log{2\pi}
\end{equation}
where $K = \begin{bmatrix}
K_{\x \x} & K_{\x v}\\
K_{v\x} & K_{vv}
\end{bmatrix}$.

The log marginal likelihood function \eqref{eq:log_likelihood} contains the unknown model parameters $\Theta$ and the kernel hyper-parameters $\bm{\beta}$.
The optimal model parameters $\Theta^*$ and hyper-parameters $\bm{\beta}^*$ can be obtained simultaneously by maximizing the log marginal likelihood function \eqref{eq:log_likelihood}, which verifies that the proposed method is a "one-step" technique for the parameter estimations and it shows that there is no need to  explicitly compute or estimate the dynamical system solutions $\x$.
An attractive property of the proposed method lies on the accessible gradient information which allows us to employ different efficient optimization algorithms.
Based on equation \eqref{eq:log_likelihood}, We can easily obtain the partial derivatives of the marginal likelihood w.r.t the model parameters $\frac{\partial }{\partial \theta_i} \log p([Y,\v]^\top|\mathcal{D},\mathcal{D}^v,\Theta,\Beta) $ and the kernel hyper-parameters $\frac{\partial }{\partial \beta_i} \log p([Y,\v]^\top|\mathcal{D},\mathcal{D}^v,\Theta,\Beta) $, respectively.
These gradient information could speed up the optimization procedure by using gradient-based optimization algorithm. 

\begin{remark}
Semi-ADAM: Due to the non-parameter model of Gaussian process, we must use all data in $\mathcal{D}$ for training. 
But for the model constraint $\mathcal{D}^v$, which should be satisfied in the whole time interval, we can randomly select parts of data for training.
The batch size can be $2n$ which contains $n$ fixed observations data and $n$ random constraint data sampled from $\mathcal{D}^v$ which are specified in Section \ref{set:samples}.
Here we employ a variation of SGD: ADAM for optimization\cite{kingma2014adam}.
\end{remark}

\subsection{Samples for model constraint}
\label{set:samples}
Theoretically, the model constraint should be satisfied through the time interval of the dynamical system and then the time indexes for model constraint $v$ could be sampled uniformly in the interval $[0, t_{max}]$.
However, it is a recognized reality that the precision of any estimator for $\x(t_i)$ diminishes with the temporal gap between $t_i$ and the observed time index provided, leading to an escalation in the uncertainty of the estimation $\x$ particularly in scenarios characterized by sparse and/or non-uniform data\cite{lee2023learning}.
Therefore, it is prudent to appropriately lift the constraint at a time significantly distant from the observed index.
Here we propose to sample $n_c$ time indexes, stored in $\mathcal{D}^v$, for numerical computation in Equation \eqref{eq:log_likelihood} and \eqref{eq:inference} to handle these special cases.

The samples for the model constraint $v$ can be obtained by MCMC method and the target distribution is formulated as $q(t)$,
\begin{equation}
q(t)\sim \exp(-\frac{V(t)}{c}),
\end{equation}
where $c$ is a constant. 
$V(t)$ is the potential function and here we define it as the posterior variance estimator with the given data positions $T_d$, i.e.,
\begin{equation}
\label{eq:constraint_sample}
V(t) = K_{vv}^{\Theta}(t,t)-K_{v\x}^{\Theta}(t,T_d)K_{\x \x}^{\Theta}(T_d,T_d)^{-1}K_{\x v}^{\Theta}(T_d,t),
\end{equation}
where $T_d$ is the data locations.
Based on the Gaussian process assumption of $v$, we known $V(t)\in [\sigma^2_v,  \eta]$, where $\sigma^2_v$ is the preset small value for regularization and $\eta$ can be computed from the specific kernel expression $k_{vv}$. 
It should be noted that the hyper-parameters $\Beta$ and model parameters $\Theta$ used in \eqref{eq:constraint_sample} are given with the optimization.
So an efficient sampling algorithm for $T_c$ is needed.
 
Due to $q(t)$ is a multimodal distribution, MH-sampler method would be difficulty and inefficient in this scenario.
Here we propose to use a simple rejection sampling algorithm for sampling the time indexes for $\mathcal{D}^v$.
The algorithm is listed below,

\begin{algorithm}[H]
\label{alg:sampling}
\caption{ Rejection sampling}
\SetKwInOut{Input}{Input:}
\SetKwInOut{Output}{Output:}
\SetAlgoLined
\Input{number of samples for constraint $n_c$, $V(t)$}
\Output{time indexes $T_c = \{ t_i\}_{i=1,\dots,n_c}$ }
$i=1$\;
\While{$i<n_c$}
{$t_i \sim \mathcal{U}(0, t_c)$ and $u \sim \mathcal{U} (0,1)$\;
$q(t_i) = \exp (-\frac{V(t_i)-\sigma^2_{\nu}}{1/4(\eta-\sigma^2_{\nu})})$\;
\If{$u < q(t_i)$}
{accept $t_i$ and $i = i+1$\;}}
\end{algorithm}

Vector arithmetic and parallel computing techniques can be used in Algorithm \ref{alg:sampling} for speeding up the sampling process.
We would like to emphasize that this sampling procedure would not be necessary for the uniform, non-sparse data scenario. It is an additive procedure for handling more challenge situations where the data locations are sparse and non-uniform.

\subsection{Prediction}
\label{sub:pred}
Once the optimal model parameter $\Theta^*$ and hyper-parameters $\Beta^*$ are obtained in Section \ref{set:inference}, additionally we could obtain the estimations of the $i$th component, $j$th order derivative of  solution, $\hat{x}^{(j)}_i,i=1,\dots,D,j=0,\dots,M$.
The posteriors of $\hat{\x}^{(j)}$ are computed in Bayesian framework which similar to the standard Gaussian process regression and their the mean functions would satisfy  the $D-1$ model constraints (except for $k$th Equation in \eqref{eq:linear_dynamic_g} ) theoretically and the $k$th equation constraint in numerically. 

For notation simplicity, we denote 
\begin{equation}
    \mathcal{X} = [Y,v]^\top\quad\text{and} \quad
    K_{\mathcal{X},\mathcal{X}} = \begin{bmatrix}
    K_{\x \x} & K_{\x v}\\
    K_{v \x} & K_{vv}
\end{bmatrix}.
\end{equation}
Then we have the joint distribution for $Y$, $v$ and $x^{(j)}_i$:
\begin{equation}
\label{eq:inference}
    \begin{bmatrix}
    \mathcal{X}\\
    x^{(j)}_i
    \end{bmatrix}
    \sim 
    \mathcal{N}(
    \begin{bmatrix}
    \textbf{0}\\
    0
    \end{bmatrix}
    ,
    \begin{bmatrix}
    K_{\mathcal{X}\mathcal{X}} & K_{\mathcal{X}x^{(j)}_i}\\
    K_{x^{(j)}_i\mathcal{X}} & K_{x^{(j)}_i x^{(j)}_i}
    \end{bmatrix}).
\end{equation}
Based on the Bayesian formula $\pi(x^{(j)}_i|\mathcal{X}) = \frac{\pi(x^{(j)}_i , \mathcal{X})}{\pi(\mathcal{X})}$, we can get the posterior distribution 
\begin{equation}
\label{eq:poterior}
    \pi(x^{(j)}_i(T_*)|T,\mathcal{X}) \sim \mathcal{N}(\bm{\mu}(x^{(j)}_i(T_*)), \Sigma(x^{(j)}_i(T_*))),
\end{equation}
where $T = [T_d, T_c]$, $T_*$ stacks the estimated time indexes and 
\begin{equation}
\label{eq:posterior_detail}
\begin{split}
        \bm{\mu}(x^{(j)}_i(T_*)) &= K_{x^{(j)}_i \mathcal{X}}(T_*,T)[K_{\mathcal{X},\mathcal{X}}(T,T)]^{-1} Y,\\
        \Sigma(x^{(j)}_i(T_*))) &=K_{x^{(j)}_i x^{(j)}_i}(T_*,T_*) - K_{x^{(j)}_i \mathcal{X}}(T_*,T)[K_{\mathcal{X}\mathcal{X}}(T,T)]^{-1} K_{\mathcal{X}x^{(j)}_i}(T,T_*).
\end{split}
\end{equation}
It is noteworthy to highlight that the equations \eqref{eq:poterior} and \eqref{eq:posterior_detail} yield the posterior estimation of $x^{(j)}_i$, and the mean functions of these posteriors not only precisely adhere to the model constraints \eqref{eq:linear_dynamic_g} but also align with the principles of physical derivation.

\subsection{Convergence}
Basic Gaussian process regression method is one type of kernel approach which are constructed in Reproducing Kernel Hilbert Space (RKHS).
The posterior mean for GP regression can be rewritten as the function which minimizes the functional \cite{williams2006gaussian}. 
\begin{equation}
    \label{eq:functional_express}
    J[u]=\frac{1}{2}\sum^D_{j=1}||x_j||^2_{\mathcal{H}_j} + \frac{1}{2}||v||^2_{\mathcal{H}_v} + \sum_{j=1}^D\frac{1}{2\sigma^2_{y_j}}\sum^n_{i=1}(y_{j,i} - x_j(t_i))^2 +  \frac{1}{2\sigma^2_{v}}\sum^{n_c}_{i=1}(0 - v(t_i))^2
\end{equation}
where  $||x_j||^2_{\mathcal{H}_j}$ is the RKHS norm corresponding to kernel $k_{x_j,x_j}$. $y_{j,i}$ means  the $j$th component of observation $\y$ in time index $t_i$.
As mentioned above $\mathcal{L}^\Theta_j u =  x_j$ and $\mathcal{L}^\Theta_v u = v$.
Because the parameters $\Theta$ can be arbitrary and would not affect the convergence,
we delete the notation $\Theta$ in $\mathcal{L}$ for simplicity.
The equation \eqref{eq:functional_express} obviously make sense in penalty term which includes the data term and regularization term.
Our goal is to understand the behaviours of this solution as $n \to \infty$.

Let $\mu(t,\y)$ be the probability measure from which the data pairs $(t_i,\y_i)$ are generated.
Observe that for $j$th component,
\begin{equation}
    \mathbb{E}\big[\sum^n_i(y_{j,i} -  x_j(t_i) \big]=n\int (y_j - x_j(t))^2d\mu(t,y_j).
\end{equation}
Let $\mathcal{L}_j\eta(t)=\mathbb{E}[y_j|t]$ be the regression function corresponding to the probability measure $\mu$ and we assumed that $\eta(t)$ is sufficiently well-bahaved so that it can be represented by the generalized Fourier series $\sum^\infty_{i=1}\eta_i\phi_i(t)$. 
The variance around $\mathcal{L}_j\eta(t)$ is denoted $\sigma^2_{y_j} =\int (y_j - \mathcal{L}_j\eta(t))^2d\mu(y_j|t)$.
Then writing $y_j - x_j = (y_j - \mathcal{L}_j\eta) + (\mathcal{L}_j\eta - x_j)$ we obtain
\begin{equation}
\label{eq:y_f_sigma}
    \int (y_j - x_j(t))^2d\mu(t,y_j) = \int (\mathcal{L}_j\eta(t) - \mathcal{L}_ju(t))^2d\mu(t) +\int \sigma^2_{y_j}d\mu(t),
\end{equation}
as the cross term vanishes due to the definition of $\mathcal{L}_j\eta(x)$.

As the second term on the right hand side of Equation \eqref{eq:y_f_sigma} is independent of $\mathcal{L}_ju$, an idealization of the regression problem consists of minimizing the functional 
\begin{equation}
\label{eq:J_functional}
    J_\mu[u] =\sum^D_{j=1}\frac{n}{2\sigma^2_{y_j}}\int (\mathcal{L}_j\eta(t) - \mathcal{L}_j u(t))^2d\mu(t) 
    + \frac{1}{2}\sum^D_{j=1}||\mathcal{L}_j u||^2_{\mathcal{H}_j} + \frac{1}{2}(\frac{n_c}{\sigma^2_v}+1)||\mathcal{L}_v u||^2_{\mathcal{H}_v}
\end{equation}
 In equation \eqref{eq:J_functional}, we can find that the number of numerical constraint points, $n_c$, is actual an weight value for the regularization term $||\mathcal{L}_v u||^2_{\mathcal{H}_v}$.
The form of the minimizing solution is most easily understood in terms of the eigenfunctions $\{\phi_i(t)\}$ of the kernel $k_{uu}$ w.r.t. to $\mu(t)$, where $\int \phi_i(t)\phi_j(t)d\mu(t) = \delta_{ij}$, see section 4.3 in \cite{williams2006gaussian}.

$$u(t) = \sum^{\infty}_{i=1}f_i\phi_i(t)$$,
and 
\begin{equation}
\begin{split}
    \mathcal{L}_j [u(t)] 
    &= \sum^{\infty}_{i=1}f_i \mathcal{L}_j[\phi_i(t)]\\ 
    &=\sum^\infty_{i=1}f_i\int k_{x_j x_j}(s,t)\phi_i(s)ds\\
    &= \sum^\infty_{i=1} f_i \lambda^j_i \phi_i(t).
\end{split}
\end{equation}
Similarly, we have $\mathcal{L}_v [u(t)]= \sum^\infty_{i=1} f_i \lambda^v_i \phi_i(t)$.
The kernels $k_{x_j x_j}(\cdot,\cdot)$, $j=1,\dots,D$ and $k_{v v}(\cdot,\cdot)$ are covariance functions corresponding to random variables $x_j$ and $v$, respectively, and derive from the nondegenerate kernel $k_{uu}(\cdot,\cdot)$. 
$\mathcal{L}$ is bound continuous linear operator in RKHS.
$\int \phi_i(t)\phi_j(t)d\mu(t)=\delta_{ij}$.
Consider a real positive semidefinite kernel $k_{x_j x_j}(t,t')$ with an eigenfunction expansion $k_{x_j x_j}(t,t')=\sum^N_{i=1}\lambda^j_i\phi_i(t)\phi_i(t')$. 
Equation \eqref{eq:J_functional} can be written as
\begin{equation}
    J_\mu[u]=\sum^D_{j=1}\frac{n}{2\sigma^2_{y_j}}\sum^\infty_{i=1}(\eta_i\lambda^j_i-f_i\lambda^j_i)^2 
    + \frac{1}{2}\sum^D_{j=1}\sum^\infty_{i=1}f_i^2\lambda^j_i + \frac{1}{2}(\frac{n_c}{\sigma^2_v}+1)\sum^\infty_{i=1}f_i^2\lambda^v_i.
\end{equation}
This is readily minimized by differential w.r.t. each $f_i$ to obtain
\begin{equation}
    f_i = \frac{1}{1+A}\eta_i,
\end{equation}
where 
\begin{equation}
        A =\frac{1}{n}\frac{\sum_{j=1}^D\lambda_i^j + (\frac{n_c}{\sigma_v^2}+1)\lambda_i^v}{\sum_{j=1}^D\frac{{\lambda_i^j}^2}{\sigma_{y_j}^2}}.
\end{equation}
Notice that the term $\frac{1}{n}
\to 0$ as $n\to \infty$ so that in this limit we would expect that $u(t)$ will converge to $\eta(t)$. 
It is worth mentioning that though our proof is based on the situation that all $D$ solutions can be observed, a part components of solutions or even only one solution can be observed in practice would be more common and the convergence result would also be existed.

\section{Linearization for nonlinear dynamical system}
\label{set:linearization}

When dealing with nonlinear dynamical systems and incomplete operating parameters, direct equation solving often proves challenging or unfeasible. Even with all parameters at hand, sophisticated computational and mathematical tools are essential for fully resolving the nonlinear ordinary differential equations required to model the process accurately. To streamline this modeling process and derive approximate functions for describing the system, researchers typically linearize the nonlinear dynamical systems and utilize matrix mathematics to solve the resultant linearized equations.

In order to linearize an nonlinear dynamical system, we propose a piecewise linearization method which is based on Taylor expansion and inspired by steady state point linerization \cite{drazin1992nonlinear}.

\subsection{Taylor series expansion for nonlinear dynamical system}
For the $i$th equation in dynamical system \eqref{eq:dynamic_system},
\begin{equation}
   \dot{x}_i(t) = f_i(\x) . 
\end{equation}
Taylor series is a series expansion of a function about a point. 
 If given a fixed point $\s = [s_1,\dots,s_D]^\top \in \mathbb{R}^D$, an expansion of the real function $f_i(\x)$ is given by:
\begin{align}
 f_i(\x) &\approx f_i(\s) + \sum_{j=1}^{D}\frac{\partial f_i(\s)}{\partial x_j}(x_j-s_j)\\
 &\approx \sum_{j=1}^{D}\frac{\partial f_i(\s)}{\partial x_j}x_j + f_i(\s) - \sum_{j=1}^{D}\frac{\partial f_i(\s)}{\partial x_j}s_j
\end{align}

Taylor's theorem states that any function satisfying certain conditions can be expressed as a Taylor series.
For simplicity's sake, only the first two terms (the zero- and first-order) terms of this series are used in Taylor approximations for linearizing ODEs.
The variable ``$\s$'' in the Taylor series is the point chosen to linearize the function around and called {\it{fixed point}} here.

In the context of dynamical systems, the Jacobian is expressed as a matrix that encapsulates the constants required to describe the system's linear characteristics. It can be interpreted as the degree to which the system is transformed to exhibit linear behavior. The Jacobian matrix, always square in form, reveals the rate of variation of each equation with respect to each variable.
The Jacobian matrix is defined as:
\begin{equation}
J(x_1,\dots,x_D) = \begin{bmatrix}
\frac{\partial f_1}{\partial x_1} & \cdots & \frac{\partial f_1}{\partial x_D}\\
\vdots & \vdots & \vdots\\
\frac{\partial f_D}{\partial x_1} & \cdots & \frac{\partial f_D}{\partial x_D}\
\end{bmatrix},
\end{equation}
and is used as such for this nonlinear set of equations \eqref{eq:dynamic_system}.
Given fixed point pair $\{t, \s\}$. the equation \eqref{eq:dynamic_system} can be approximate linearly by
\begin{equation}
\begin{bmatrix}
\dot{x_1}\\
\vdots\\
\dot{x_D}
\end{bmatrix}
=
\begin{bmatrix}
J_{11}&\dots&J_{1D}\\
\vdots&\ddots&\vdots\\
J_{D1}&\dots&J_{DD}
\end{bmatrix}
\begin{bmatrix}
x_1\\
\vdots
\\x_2
\end{bmatrix}
+
\begin{bmatrix}
c_1\\
\vdots\\
c_D
\end{bmatrix},
\end{equation}
where $J_{ij} = \frac{d f_i}{d x_j}|_{\x = \s}$ and $c_i=f_i(\s) - \sum_{j=1}^{D}\frac{\partial f_i(\s)}{\partial x_j}s_j$.

\subsection{Piecewise linearization}
linearization on one fixed point is obviously a bad approximation for $f_i(\x(t))$ in the whole time interval. Here we propose to take a series of points $\{t_k, \s_k \}, k=1,\dots,n$ as fixed points and approximate the function $f_i(\x)$ by a piecewise function $\tilde{J}_{ij}(t)$.
Then we can define some piecewise functions as following
\begin{equation}
\begin{split}
\tilde{J}_{ij}(t)&=\frac{d f_i}{d x_j}|_{\x = \s_k}\\
\tilde{c}_i(t)&=f_i(\x) - \sum_{j=1}^{D}\frac{d f_i(\x)}{d x_j}x_j |_{\x = \s_k},
\end{split}
\end{equation}
where $k = \argmin_{k}(|t - t_k|)$.
Then the nonlinear differential equation \eqref{eq:dynamic_system} can be approximated by the following linear equations 
\begin{equation}
        \dot{\x} = \tilde{\bm{J}}(t) \x+ \tilde{\bm{c}}(t).
\end{equation}

For large noisy observations, we need infer these fixed points from data and obtain the posterior distribution $p(\s|Y)$ which can be estimated by standard Gaussian process regression.
For this situation, we give a modified expression \eqref{eq:log_likelihood}
\begin{equation}
\begin{split}
\label{eq:log_marginal_post_s}
    \log p([Y,\v]^\top|\mathcal{D},\mathcal{D}^v,\Theta,\Beta) &= \int \log p([Y,\v]^\top|\mathcal{D},\mathcal{D}^v,\Theta,\Beta,\s) p(\s|Y) d\s \\
    &=\mathbb{E}_{\s\sim p(\s|Y)}\log p([Y,\v]^\top|\mathcal{D},\mathcal{D}^v,\Theta,\Beta,\s) ,
    \end{split}
\end{equation}
for the marginal likelihood function \eqref{eq:log_likelihood}.
Usually the uncertainty of $p(\s|Y)$ would be small because $Y$ is the noisy measurement of $\s$ and this integration can be computed by Monte Carlo method.

If the observations has no or small noise, we can directly use the observations or its interpolation values as the fixed points. 
Then Equation \eqref{eq:log_marginal_post_s} is simplified as 
\begin{equation}
    \log p([Y,\v]^\top|\mathcal{D},\mathcal{D}^v,\Theta,\Beta) =  \log p([Y,\v]^\top|\mathcal{D},\mathcal{D}^v,\Theta,\Beta,\hat{\s}),
\end{equation}
 where $\hat{\s} = Y$,  indicating that $p(\s|Y)$ is represented by a Dirac delta function.
 
\section{Numerical examples}
\label{set:num_examples}
In our numerical illustrations, we apply our methodology to various dynamical systems, including both linear and nonlinear types, as well as ordinary differential equations.
First, we present an example involving a linear dynamical system to demonstrate our approach in different settings, with varying data sizes and noise levels.
Next, we showcase a scenario featuring a nonlinear second-order ordinary differential equation, highlighting the proposed method's effectiveness in higher-order nonlinear contexts. This example includes predictions of the solution and its derivatives, illustrating the accuracy of the proposed predictions.
Finally, we examine the competitive FitzHugh-Nagumo equations, comparing the proposed method with others. This exploration further underscores the versatility and applicability of the proposed methodology.

\subsection{Linear dynamical system}
\begin{figure}[h]	
    \centering
    \includegraphics[width=0.6\linewidth]{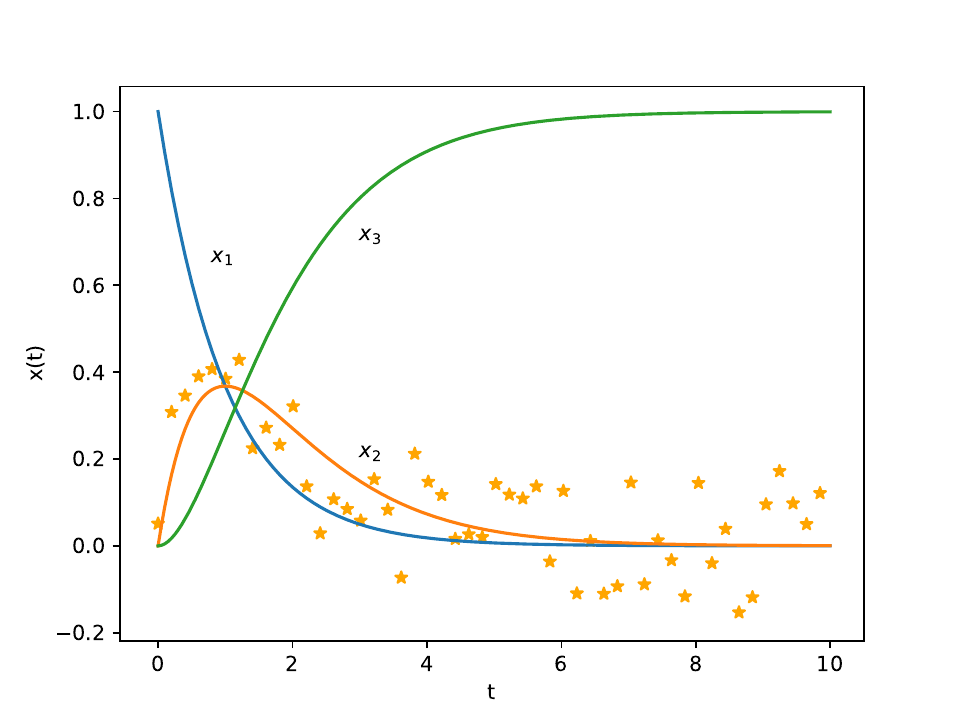}  
    \caption{The observations (stars) and solution of the linear dynamical system (solid lines).}
    \label{fig:linearode}
\end{figure}

A linear dynamical system composed of a series of reactions $x_1\stackrel{\theta_1,\theta_2}{\longrightarrow}x_2\stackrel{\theta_2}{\longrightarrow}x_3$ is considered to illustrate the properties of the proposed method.
\begin{equation}
\label{eq:examp1_ode}
    \dot{\x} = A \x,
\end{equation}
where 
\begin{equation}
    A=\begin{bmatrix}
    -\theta_1 & 0 & 0\\
    \theta_1 & -\theta_2 & 0\\
    0 &\theta_2 & 0
    \end{bmatrix}
\end{equation}
and $\theta_1 $, $\theta_2$ are two model parameters needed to be estimated and the initial condition is $x(0)=[1,0,0]^T$. The true solution of the  system is shown in  Fig \ref{fig:linearode}.

Here, we define the variable $x_2$ as $u$.
In practical situations, achieving complete observability of all variables is frequently unattainable. To evaluate the robustness of our methodology in managing partial observations, we propose a scenario wherein only observations related to $x_2$ are accessible. Consequently, the algorithm initiates by assuming a Gaussian process as the prior distribution for the variable $u$,
\begin{equation}
u:=x_2 \sim\mathcal{GP}(0, k_{uu}(t,t')).
\end{equation}
Then the system constraint discussed in Section \ref{set:Gaussian_process} is expressed as
\begin{equation}
v=u''+(\theta_1+\theta_2)u'+\theta_1\theta_2u=0,
\end{equation} 
which is also a Gaussian process,
\begin{equation}
v \sim\mathcal{GP}(0, k_{vv}(t,t')).
\end{equation}
Specifically, the corresponding kernel of constraint $v$ can be specified as 
\begin{equation}
\begin{aligned}
k_{vv} &=k_{u''u''}+(\theta_1+\theta_2)k_{u''u'}+\theta_1\theta_2k_{u''u} \\ 
  &+(\theta_1+\theta_2)k_{u'u''}+  (\theta_1+\theta_2)^2k_{u'u'}+ (\theta_1+\theta_2)\theta_1\theta_2k_{u'u} \\
 &+\theta_1\theta_2k_{uu''}+(\theta_1+\theta_2)\theta_1\theta_2k_{uu'}+\theta_1^2\theta_2^2k_{uu}.
\end{aligned}
\end{equation}

Consider a scenario involving a set of noisy data denoted as $\bm{y}$, associated with the variable $\bm{x}_2$. This noisy data is generated according to the relationship $\bm{y} = \bm{x}_2(T) + \bm{\epsilon}$, where $T$ consists of $n$ data points uniformly selected from the interval $[0, 10]$, as illustrated in Figure \ref{fig:linearode}.
The term $\bm{\epsilon}$ represents the noise, stochastically generated following a normal distribution $\mathcal{N}(\bm{0}, \sigma_n^2I_n)$ and the function $x_2(t)$ is generated by a dynamical system solver using the parameters $\bm{\theta}^*=[\theta_1^*, \theta_2^*]=[1,1]$, representing the ground truth.
Given the uniform distribution of observations, there is no need to acquire constraint points through the rejection sampling algorithm; instead, they can be directly drawn from the uniform distribution within the interval $[0, 10]$. To comprehensively assess the performance of the methodology, we compute the mean and standard deviation (SD) of the parameter estimates across 100 repeated trials.

We conduct an comparative analysis of estimations using distinct dataset sizes, namely $n = 50$ and $n = 100$, in conjunction with varying levels of noise standard deviations ($\sigma_n = 0.01, 0.05, 0.1$). 
Here and below, each experiment was repeated $100$ times for achieving the estimate mean and standard variance. The outcomes across various sample sizes and noise intensities are consolidated within Table \ref{tab:linearode}. Evidently, the proposed technique yields minimal bias and standard deviations, particularly noteworthy in the scenario of a sample size of $100$ combined with a noise level of $0.01$. 
The derived parameter estimates, $\hat{k}_1 = 0.951 \pm 0.0167$ and $\hat{k}_2 = 1.027 \pm 0.02$, closely approximate the true values.
Even in circumstances characterized by limited sample sizes and substantial observation errors, specifically when $n = 50$ and $\sigma = 0.1$,  the ground truth is still contained within the interval $[\text{Mean} - 2*\text{SD}, \text{Mean} + 2*\text{SD}]$.
However, introducing random, large observation error amplifies the variance of parameter estimates and biases mean estimates.
Furthermore, a discernible pattern emerges: the standard deviation of the estimates diminishes progressively as the observation noise level decreases, assuming a constant sample size. 
In this table, although the mean estimate of $0.918$ with $n=100$ is lower than the mean estimate of $0.972$ with $n=50$ in the highest noise level case, the standard deviation decreases to a more acceptable value. This confirms that the proposed approach yields more reasonable estimates (mean and standard deviation) as the dataset size increases, especially in scenarios with high noise levels.
In summation, the simulation outcomes robustly demonstrate the accuracy of the proposed method in inferring parameters within dynamical systems that encompass partial observations.

\setlength{\tabcolsep}{15pt}
\begin{table}\small 
	\caption{ For linear dynamical system, the metric results for different sample sizes and noise levels.} 
	\label{tab:linearode} 
	\centering 
	\begin{tabular}{c c c c c c c c } 
		\toprule 
		
 & & \multicolumn{2}{c}{$\sigma=0.01$} & \multicolumn{2}{c}{$\sigma=0.05$} &\multicolumn{2}{c}{$\sigma=0.1$}\\
 & & $k_1$ & $k_2$ & $k_1$  & $k_2$&  $k_1$ &$k_2$ \\ 
		\midrule 
$n=50$&		 Mean &0.970 & 0.991 & 0.859&0.851&0.768&0.972\\ 
	&	 SD & 0.013 & 0.021 & 0.054 &0.062 &0.150 &0.311\\	
		 \midrule 
$n=100$&		 Mean &0.951& 1.027 &0.890& 0.943& 0.853& 0.918\\ 
	&	 SD & 0.017 &0.020& 0.040 &0.050 & 0.085 & 0.100\\	
		\bottomrule 
	\end{tabular}
\end{table}

\subsection{Van der Pol euqation}
The Van der Pol equation is a typical nonlinear ODE which can generate a shock wave solution \cite{guckenheimer1980dynamics}. 
It is defiend as
\begin{equation}
\label{eq:example_van}
    u'' - \theta(1-u^2)u'+u=0,
\end{equation}
where $\theta$ is unknown parameter and the initial condition are $u(0)=2, \dot u(t)=0$. 
It can also be rewritten as a nonlinear dynamical system
\begin{equation}
\left\{
\begin{split}
    \dot{u} &= w \\
    \dot{w} &= \theta(1-u^2)w - u,
\end{split} 
\right.
\end{equation}
where $w$ represents the first order derivative of $u$.
We have observed $100$ noisy solutions, each associated with the true parameter value $\theta^* = 0.5$ and a noise level of $0.1$. Due to the compact and continuous nature of the observations, the constraint points are not selected using the sampling method. Instead, the steady-state points, denoted as ${t^k, s^k}_{k=1,\dots,n}$, are determined through conventional Gaussian Process Regression (GPR), as outlined in Algorithm \ref{alg:sampling}.
Then, a general piecewise linearization function is obtained 
\begin{equation}
\begin{bmatrix}
\dot{u}\\
\dot{w}
\end{bmatrix}
=
\begin{bmatrix}
\widetilde{J}_{11} &\widetilde{J}_{12}\\
\widetilde{J}_{21} &\widetilde{J}_{22}
\end{bmatrix}
\begin{bmatrix}
u \\
w
\end{bmatrix}
+
\begin{bmatrix}
\widetilde{c}_1(t)\\
\widetilde{c}_2(t)
\end{bmatrix},
\end{equation}
where $\widetilde{J}_{11} =0$, 
$\widetilde{J}_{12}=1$,
$\widetilde{J}_{21} =-2\theta u(t)w(t)-1$, $\widetilde{J}_{22} =\theta(1-u(t)^2)$ and $\widetilde{c}_1(t)=0$, $\widetilde{c}_2(t)=2\theta u(t)^2w(t)$.
The equation constraint is expressed as a function with respect to $u$ and its derivatives, 
\begin{equation}
v=u''-\widetilde{J}_{22}(t)u'+\widetilde{J}_{21}u-\widetilde{c}_2(t).
\end{equation}

The experiment is conducted 100 times with different dataset sizes ($n=50$ and $n=100$). Table \ref{tab: van der pol} presents the summary statistics of the parameter estimates. The results indicate that the method effectively handles nonlinear differential equations, providing estimates with high accuracy that are very close to the ground truth. The mean estimates approach the true value of $0.5$, and the standard deviation decreases as the dataset size increases.

\setlength{\tabcolsep}{15pt}
\begin{table}[H]
\small 
	\caption{ For Van der Pol equation, the metric results for different sample sizes ($\sigma=0.1$).} 
	\label{tab: van der pol} 
	\centering 
	\begin{tabular}{c c c c  } 
		\toprule 
 $n=50$ &Mean &  & 0.445 \\
  & SD &   &0.086 \\
	$n=100$ &Mean & & 0.447 \\
	& SD &   &0.060 \\		 	 	
		\bottomrule 
	\end{tabular}
\end{table}

\setlength{\tabcolsep}{15pt}
\begin{table}[H]
\small 
	\caption{ For Van der Pol equation, MSE of the state, first derivative and second derivative by the proposed method, ODE ($\theta=0.447$) and GPR, respectively ($n=100$). } 
	\label{tab: example2_MSE} 
	\centering 
	\begin{tabular}{c c c  c} 
		\toprule 
& Proposed method & ODE solver  & GPR\\
 $u$ & $\bm{0.0008}$ & 0.0053 & 0.0038 \\
 $u'$ & $\bm{0.0023}$ &0.0072 & $0.0376$ \\
 $u''$& $\bm{0.0130}$ & 0.0239 & $0.6275$ \\	 	 	
		\bottomrule 
	\end{tabular}
\end{table}

\begin{figure}[H]	
    \centering
    \includegraphics[width=\linewidth]{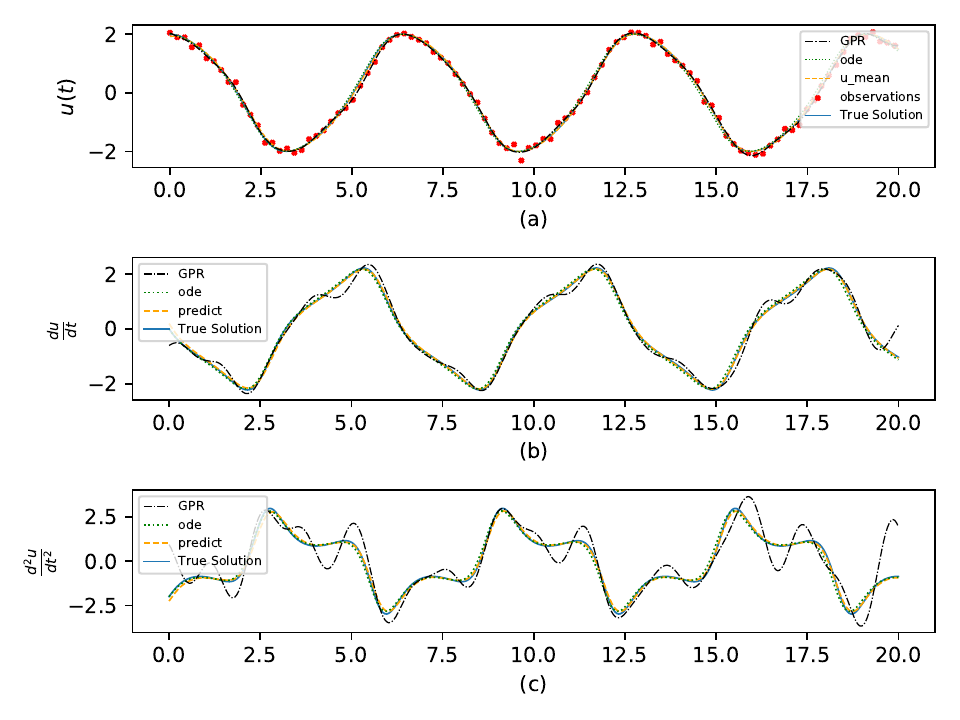}  
    \caption{Illustration of the prediction of $u, \frac{du}{dt},\frac{d^2u}{dt^2}$  using the proposed method with estimated model parameters $\hat{\theta} = 0.447$.  The blue solid curve represents the true solution and its derivative (ODE solution with $\theta=0.5$).
The red solid dots represent the  noisy observations. 
The orange dashed lines represent the estimations of solution and its derivatives.
The blue dotdash lines represent estimations of solution and its derivatives by GPR.
The green dotted lines represent the solution and its derivatives computed by DS solver with $\theta=0.447$.}
    \label{fig:pre uhg}
\end{figure}

Additionally, we use these statistical metrics not only to evaluate the quality of the estimation but also to predict the posterior mean of the state function and its derivatives at the estimated parameter values. This prediction is carried out using the constraint informed Gaussian process method as described in equation \eqref{eq:posterior_detail}.
As depicted in Figure \ref{fig:pre uhg}, the figures illustrate the posterior means predicted by the parameter values obtained during a single execution of the proposed algorithm. 
The findings reveal a favorable alignment between the predicted curves of the three functions, namely $u$, $u'$, and $u''$, and their corresponding true curves subsequent to addressing linearity in the differential equation and effectively learning the associated parameters through the proposed methodology. 
This alignment serves to substantiate the method's parameter estimation accuracy,  underscored by its robust capability to rectify the influence of noisy observations. 
This proficiency is particularly pronounced when the underlying equations are duly taken into account.

As illustrated in Figure \ref{fig:pre uhg}, the figures show the posterior means predicted by the proposed method using estimated parameter values. 
The results indicate a strong alignment between the predicted curves of the functions $u$, $\frac{du}{dt}$, and $\frac{d^2u}{dt^2}$ and their ground truths. 
Especially, the curve of predicted second-order derivative function is pretty closed to the ground-truth compared with the standard Gaussian process regression which demonstrate the important influence of constraint information in modeling and prediction.
This alignment demonstrates the method's accuracy in the estimation of high-order derivative functions, showcasing its robust ability to mitigate the effects of noisy observations, particularly when the underlying equations are properly considered.
In table \ref{tab: example2_MSE}, we can find that the estimation of DG-GPR are more accurate than the standard GPR and ODE solution with the estimated parameter $\hat{\theta}=0.447$.
Also we can find that ODE solution with the estimated parameter $\hat{\theta}=0.447$ are pretty closed to the true ODE solution ($\hat{\theta}=0.5$).

\subsection{FitzHugh-Nagumo equations}
\label{set:example3}
\begin{figure}[htbp]	
    \centering
    \includegraphics[width=\linewidth]{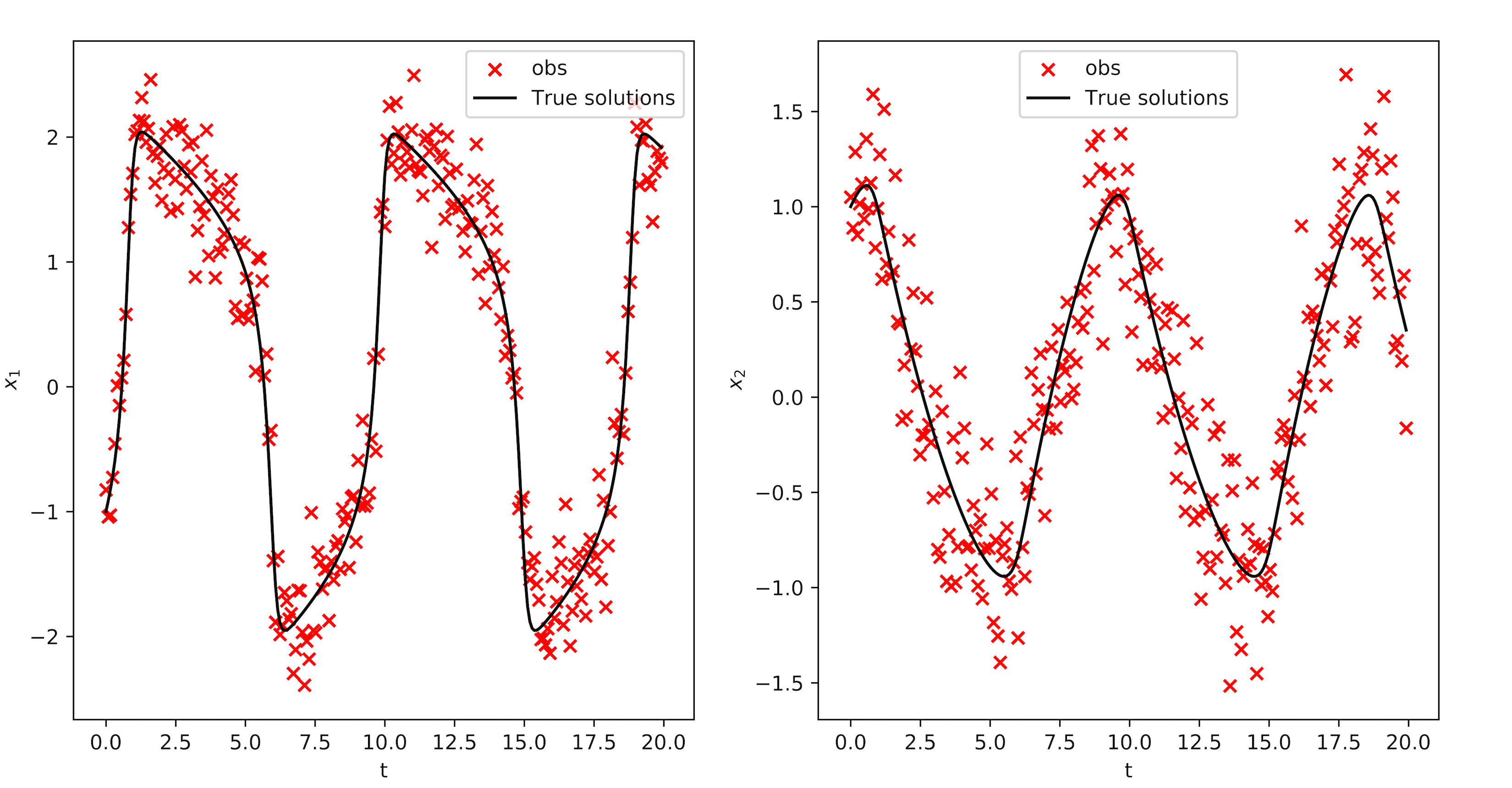}  
    \caption{True solution of the  FitzHugh-Nagumo equations and  500 simulated samples of data generated from the equations}
    \label{fig:F_N nolinear dynamical system}
\end{figure}

We also experimented with data drawn from the well-known FitzHugh-Nagumo equations \cite{ramsay2007parameter,hall2014quick}. This is a two-dimensional nonlinear dynamical system originally developed to simulate the behavior of spiking potentials in giant axons of squid neurons. The model could provide a good test for the proposed parameter inference approach. This set of equations involves three parameters and has the form
\begin{equation}
    \begin{split}
        \dot{x}_1(t) &= \theta_3\{x_1(t) - \frac{1}{3}x_1(t)^3 + x_2(t) \},\\
        \dot{x}_2(t) &= -\theta_3^{-1}\{x_1(t) - \theta_1 + \theta_2 x_2(t) \}.
    \end{split}
\end{equation}
Figure \ref{fig:F_N nolinear dynamical system} shows the numerical solution of the model for the true parameters $\theta=\{0.2,0.2,3\}$, initial values $x_1(0)=-1$, $ x_2(0)=1$  and $250$ observations with noise.
We generated data from the noisy equations
\begin{equation}
   \begin{split}
    y_1 &= x_1(t) + \epsilon_1, \\
    y_2 &= x_2(t) + \epsilon_2
   \end{split}
\end{equation}  
where $\epsilon_i \sim \mathcal{N}(0,0.3^2),i=1,2$.
In the third set of simulations, the data were chosen to have a large noise level, which highlights the ability of the method to handle highly noisy observations. The piecewise linearization function is written,
\begin{equation}
\begin{split}
\dot{x_1}(t)&= \widetilde{J}_{11}(t) x_1(t)+\widetilde{J}_{12}(t)x_2(t)+\widetilde{c}_1(t),\\
\dot{x_2}(t)&= \widetilde{J}_{21}(t) x_1(t)+\widetilde{J}_{22}(t)x_2(t)+\widetilde{c}_2(t),
\end{split}
\end{equation}
where 
\begin{equation}
\begin{split}
\widetilde{J}_{11}& =\theta_3(1-x_1(t)^2),\\ 
\widetilde{J}_{12}&=\theta_3,\\
\widetilde{J}_{21} &=-\theta_3^{-1}, \\
\widetilde{J}_{22} &=-\theta_2 \theta_3^{-1},
\end{split}
\end{equation}
and $\widetilde{c}_1(t)=\frac{2}{3}\theta_3x_1(t)^3$, $\widetilde{c}_2(t)=- \theta_1\theta_3^{-1}$.
Then $x_2(t)$ is represented as a function of $x_1(t)$,
\begin{equation}
x_2(t)=\frac{x_1(t)'-\widetilde{J}_{11}(t)x_1- \widetilde{c}_1(t)}{\widetilde{J}_{12}(t)}.
\end{equation}

The experiment was conducted 100 times, yielding mean values and standard deviations for the parameter estimates as follows: $\text{Mean} = {0.1998, 0.2511, 2.7465}$ and $\text{SD} = [0.016, 0.053, 0.090]$, as presented in Table \ref{tab: F-N equations}. The estimated mean of the first parameter closely aligns with the true value, with a negligible deviation of $0.0002$, demonstrating high accuracy. With a standard deviation of $0.016$, this parameter also exhibits low variability, reflecting both high accuracy and precision.
However, the higher standard deviations observed in the second and third parameters suggest that noise or data uncertainty may be affecting the stability of these estimates. The results indicate good accuracy for the first parameter, moderate overestimation for the second, and underestimation for the third, with varying levels of precision as reflected in their respective standard deviations.
Overall, the proposed method demonstrates the capability to produce highly precise estimates while providing reasonable variability measures for the parameters.

\setlength{\tabcolsep}{22pt}
\begin{table}[htpb]
\small 
	\caption{ For FitzHugh-Nagumo equations, the metric results with  noise level $\epsilon=0.3$.} 
	\label{tab: F-N equations} 
	\centering 
	\begin{tabular}{c c c c } 
		\toprule 
    & $\theta_1$&$\theta_2$ &$\theta_3$ \\ 
     \midrule 
 True value & $0.2$  & $0.2$ &$3.0$ \\ 
 Mean  & $0.1998$  & $0.2511$ &$2.7465$\\
  SD & $0.016$  & $0.053$& $0.090$ \\ 	 	
		\bottomrule 
	\end{tabular}
\end{table}

We also compared the proposed approach with the method of Hall and Ma (2014) \cite{hall2014quick}, under the same conditions (the true value of $\theta$ was $(5, 1, 0.5)^T$, and sample sizes $n=250$, with standard deviations $\sigma = 0.1$, $\sigma = 0.2$, $\sigma = 0.3$). The results are presented in Table \ref{tab:FN_equation}.
Our findings indicate that the mean estimates of $\theta_1$ and $\theta_2$ are more accurate than those obtained using the compared method across all noise levels. However, for the estimation of $\theta_3$, the mean estimates obtained via the proposed method did not perform as well. 
When comparing the standard deviations, our proposed method provides values comparable to the compared method for $\theta_2$ and $\theta_3$. For the estimation of $\theta_1$, the proposed method delivers a more accurate estimate, while the compared method yields an excessively large value.

\setlength{\tabcolsep}{6pt}
\begin{table}\small 
	\caption{ For FitzHugh-Nagumo equations, the metric results for different sample sizes and noise levels.} 
	\label{tab:FN_equation} 
	\centering 
	\begin{tabular}{c c c c c c c c c c c} 
		\toprule 
		
  &\multicolumn{2}{c}{$\sigma=0.1$} & &\multicolumn{2}{c}{$\sigma=0.2$} & &\multicolumn{2}{c}{$\sigma=0.3$}\\
  & $\theta_1$ & $\theta_2$ & $\theta_3$ & $\theta_1$ & $\theta_2$ & $\theta_3$ & $\theta_1$ & $\theta_2$ & $\theta_3$ \\ 
		\midrule 
Proposed method \\  
   Mean& $\bm{4.9987}$ &$\bm{0.9997}$&$0.4781$&$\bm{4.9954}$&$\bm{1.0006}$&$0.4671$&$\bm{4.9275}$&$\bm{0.9723}$&$0.4716$\\ 
SD&$0.0527$&$0.0201$&$0.0088$&$0.1173$&$0.0450$&$0.0163$&$0.1766$&$0.0680$&  $0.0278$\\ 

Compared method \cite{hall2014quick} \\
   Mean&4.9363 &$0.9851$&$\bm{0.4949}$&$4.9222$&$0.9772$&$\bm{0.4903}$&$4.7183$&$0.9392$&$\bm{0.4868}$\\ 
   SD&$2.2596$&$0.5110$&$0.0218$&$3.3960$&$0.7688$&$0.0283$&$3.8731$&$0.8806$&$0.0323$\\ 

		\bottomrule 
	\end{tabular}
\end{table}

\section{Conclusion}

This paper addresses the problem of parameter estimation in a specific class of dynamical systems, where each component of the system solution is influenced by a latent variable $u$ and its derivatives. A Model-Embedded Joint Inference Framework is introduced, integrating both data and model information. The key methodology employed is Gaussian Process Regression,  a non-parametric kernel-based method that ensures adherence to physical laws and model constraints.
To balance the fitting of finite data with the infinite model constraints, the semi-ADAM optimizer is employed for training. Additionally, the stochastic nature of the approach allows uncertainty quantification through repeated fitting, providing a measure of the reliability of the parameter estimates. The method demonstrates convergence in the linear case, while for nonlinear dynamical systems, a piecewise linearization scheme is proposed, transforming the nonlinear system into a linear approximation.

At the core of the proposed method is the construction of joint inference, combining observational data with model constraints using GPR. 
Furthermore, the method quantifies the uncertainty of the estimates, enhancing the robustness of the results.
However, a limitation lies in the complexity of deriving analytical expressions for the kernel functions corresponding to different elements of the dynamical solution. 
Additionally, the theoretical equivalence between formulation of general dynamical systems and their latent variable reformulations remains unresolved, and further investigation is needed to establish this connection.

\section*{Acknowledgement}
Thanks Ph.D Lu Lu for meaningfull discussion.
Xiang Zhou acknowledges the support from Hong Kong General Research Funds (11308121, 11318522, 11308323), and the NSFC/RGC Joint Research Scheme [RGC Project No. N-CityU102/20 and NSFC Project No. 12061160462].
Hongqiao Wang acknowledges the support from Natural Science Foundation of China (Grant No.12101615) and the Natural Science Foundation of Hunan Province (Grant No. 2022JJ40567).
This work was supported in part by the High Performance Computing Center of Central South University.

\appendix
\section*{Appendices}

\section{Kernel matrices in GPR}
\label{set:appgpr}
 
Assume $u(\x)$ is a Gaussian random field with $\dim$-dimension vector input $\x$.
The covariance relationship between $u(\x)$ and $u(\x')$ can be computed by a kernel function $k_{u,u}$.
Benefiting from the Gaussian process property, the first and second order derivatives are also Gaussian processes.
Here we denote $\b=[b_1,\dots,b_d]^\top$ as the gradient vector and $J = [J_{11}, \dots, J_{ij},\dots,J_{dd}]^\top$ as the Hessian matrix of $u$. 
The covariance functions between different random variables are denoted by 
\begin{equation*}
    K_{u,\b}= [K_{u,b_1},\cdots,K_{u,b_d}],
\\ K_{b,b} =
\begin{bmatrix}
K_{b_1,b_1}&\cdots&K_{b_1,b_\dim}\\
\vdots& &\vdots\\
K_{b_\dim,b_1}&\cdots&K_{b_\dim,b_\dim}
\end{bmatrix},
\end{equation*}
\begin{equation*}
    K_{u,J}= [K_{u,J_{11}},\cdots,K_{u,J_{ij}},\cdots,K_{u,J_{dd}}],
\end{equation*}
\begin{equation*}
 K_{\b,J} = 
 \begin{bmatrix}
K_{b_{1},J_{1,1}}& K_{b_{1},J_{1,2}} &  \cdots & K_{b_{1},J_{2,1}} &\cdots &k_{b_{1},J_{\dim,\dim}}\\
\vdots&  & \vdots& & \vdots&\\
K_{b_{\dim},J_{1,1}}& K_{b_{\dim},J_{1,2}} & \cdots& K_{b_{\dim},J_{2,1}}&  \cdots & K_{b_{\dim},J_{\dim,\dim}}
\end{bmatrix},
\end{equation*}

\begin{equation*}
 K_{J,J} = 
 \begin{bmatrix}
K_{J_{1,1},J_{1,1}}& K_{J_{1,1},J_{1,2}} &  \cdots & K_{J_{1,1},J_{2,1}} &\cdots &k_{J_{1,1},J_{\dim,\dim}}\\
K_{J_{1,2},J_{1,1}}& K_{J_{1,2},J_{1,2}} &  \cdots & K_{J_{1,2},J_{2,1}} &\cdots &K_{J_{1,2},J_{\dim,\dim}}\\
\vdots&  & \vdots& & \vdots&\\
K_{J_{\dim,\dim},J_{1,1}}& K_{J_{\dim,\dim},J_{1,2}} & \cdots& K_{J_{\dim,\dim},J_{2,1}}&  \cdots & K_{J_{\dim,\dim},J_{\dim,\dim}}
\end{bmatrix},
\end{equation*} 
where 
\begin{equation*}
\begin{split}
K_{b_i,b_j} &=\frac{\partial^2 }{\partial x_i \partial x'_j} K_{u,u}, \quad
K_{b_j,u} =\frac{\partial }{\partial x_j} K_{u,u}, \quad K_{J_{i,j},b_{e}}=\frac{\partial^2 }{\partial x_i \partial x_j } \frac{\partial }{ \partial x'_e}K_{u,u},\\
K_{J_{i,j},b_{e}} &= K_{b_{e}, J_{i,j}}^\top,\quad K_{J_{i,j},J_{e,g}}=\frac{\partial^2 }{\partial x_i \partial x_j } \frac{\partial^2 }{ \partial x'_e \partial x'_g}K_{u,u}, \quad
K_{J_{i,j}, u} =\frac{\partial^2 }{\partial x_i \partial x_j} K_{u,u}.
\end{split}
\end{equation*}

\bibliographystyle{plain}

\end{document}